\documentclass[twocolumn,showpacs,aps,prl,superscriptaddress]{revtex4}

\usepackage{graphicx}
\usepackage{dcolumn}
\usepackage{amsmath}
\usepackage{epsfig}

\input pubboard/babarsym

\newcommand{\BABARPubYear}    {05}
\newcommand{\BABARPubNumber}  {31}

\newcommand{\SLACPubNumber} {11288}

\def\figurebox#1#2#3{%
    \def\arg{#3}%
    \ifx\arg\empty
    {\hfill\vbox{\hsize#2\hrule\hbox to #2{\vrule\hfill\vbox to #1{\hsize#2\vfill}\vrule}\hrule}\hfill}%
    \else
    {\hfill\epsfbox{#3}\hfill}%
    \fi}

\renewcommand{\DeltaE}{\ensuremath{\Delta E^*}\xspace}

\newcommand{\ccc}{\ensuremath{\Kp\pim\pip\gamma}\xspace}
\newcommand{\ccn}{\ensuremath{\Kp\pim\piz\gamma}\xspace}
\newcommand{\scc}{\ensuremath{\KS\pim\pip\gamma}\xspace}
\newcommand{\scn}{\ensuremath{\KS\pip\piz\gamma}\xspace}
\newcommand{\mkpipi}{\ensuremath{m_{K\pi\pi}}\xspace}
\newcommand{\btosg}{\ensuremath{b\rightarrow s\gamma}\xspace}
\newcommand{\kpg}{\ensuremath{K\pi\gamma}\xspace}
\newcommand{\kppg}{\ensuremath{K\pi\pi\gamma}\xspace}

\begin{document}

\preprint{\babar-PUB-\BABARPubYear/\BABARPubNumber} 
\preprint{SLAC-PUB-\SLACPubNumber} 

\begin{flushleft}
\babar-PUB-\BABARPubYear/\BABARPubNumber\\
SLAC-PUB-\SLACPubNumber\\
\end{flushleft}

\title{\large\bf\boldmath
Measurement of branching fractions and mass spectra of 
$B\to K\pi\pi\gamma$
}

%
\author{B.~Aubert}
\author{R.~Barate}
\author{D.~Boutigny}
\author{F.~Couderc}
\author{Y.~Karyotakis}
\author{J.~P.~Lees}
\author{V.~Poireau}
\author{V.~Tisserand}
\author{A.~Zghiche}
\affiliation{Laboratoire de Physique des Particules, F-74941 Annecy-le-Vieux, France }
\author{E.~Grauges}
\affiliation{IFAE, Universitat Autonoma de Barcelona, E-08193 Bellaterra, Barcelona, Spain }
% ...
\collaboration{The \babar\ Collaboration}
\noaffiliation

\date{\today}

\begin{abstract}
We present a measurement of the partial branching fractions and mass
spectra of the exclusive radiative penguin processes $B\to
K\pi\pi\gamma$ in the range $\mkpipi < 1.8\gevcc$.  We reconstruct four
final states: \ccc, \ccn, \scc, and \scn, where $\KS\to\pip\pim$.  Using
232 million $\epem\to\BB$ events recorded by the \babar\ experiment at
the \pep2 asymmetric-energy storage ring, we measure the branching
fractions
$\BR(\Bp\to\ccc)
 =(2.95\pm 0.13\,(\mathrm{stat.})\pm 0.20\,(\mathrm{syst.}))\times 10^{-5}$, 
$\BR(\Bz\to\ccn)
 =(4.07\pm 0.22\,(\mathrm{stat.})\pm 0.31\,(\mathrm{syst.}))\times 10^{-5}$,
$\BR(\Bz\to K^0\pi^+\pi^-\gamma)
 =(1.85\pm 0.21\,(\mathrm{stat.})\pm 0.12\,(\mathrm{syst.}))\times 10^{-5}$, 
and
$\BR(\Bp\to K^0\pi^+\pi^0\gamma)
 =(4.56\pm 0.42\,(\mathrm{stat.})\pm 0.31\,(\mathrm{syst.}))\times 10^{-5}$.  

\end{abstract}


\maketitle


In the standard model (SM) the radiative penguin decay $B\to X_s\gamma$,
where $X_s$ is a hadronic system with unit strangeness, proceeds via
weak-interaction loop diagrams. New physics, beyond the SM, may also
contribute to the loop amplitude, and lead to differences
from the SM.  This possibility has been pursued in inclusive
measurements, which are theoretically clean but experimentally
challenging, and in exclusive measurements, such as $B\to K\pi\gamma$.
We report measurements of the branching fractions and mass spectra
for the decays $B\to K\pi\pi\gamma$ in four channels.  SM
predictions of the rates and resonance structure of these decays 
have large uncertainties~\cite{SMpred}.  The
$K^+\pi^+\pi^-\gamma$ and $K^0\pi^+\pi^-\gamma$ decay channels have
previously been observed~\cite{Belle}.  Throughout this Letter, stated
decays include charge conjugate modes.

The decays $B\to K\pip\piz\gamma$, which have not previously been
observed, are of particular interest because these three-body hadronic
states permit the measurement, given sufficient statistics, of the
photon polarization~\cite{Gronau02}. The polarization measurement
depends on the interference between processes such as
$(K\pi^+)\pi^0\gamma$ and $(K\pi^0)\pi^+\gamma$, where $()$ indicates
resonant substructure.  This measurement may be compared with the SM
prediction of nearly complete left-handed polarization.

We use a sample of $(232 \pm 1.5)\times 10^6$ \BB\ pairs 
in a 210.9 fb$^{-1}$ dataset collected at
the \FourS\ resonance with the \babar\ detector at the \pep2\
asymmetric-energy $e^+e^-$ collider.  For background studies, we also
use a $21.7\ \hbox{fb}^{-1}$ sample collected below the \BB threshold.
The measurement procedure was designed using 
simulated signal and background events,
data in sideband kinematic regions, and
reconstructed $B\to D\pip$, $D\to K\pi\pi$ decays.  Only after we
established the selection and fit procedures did we examine signal
candidates in the data sample.

A description of the detector exists
elsewhere~\cite{detector}.  For this measurement, the most important
detector elements are the five-layer silicon microstrip tracking
detector (SVT) and the forty-layer drift chamber (DCH), situated in a
1.5\,T solenoidal magnetic field, which measure charged particle
momenta; the CsI(Tl) electromagnetic calorimeter (EMC), which measures
the energies and directions of the photons; and the detector of
internally reflected Cherenkov light (DIRC). The DIRC response and
energy loss (\dedx) measured in the SVT and DCH are used to identify
charged kaons and pions.


We reconstruct the photon candidate in the $K\pi\pi\gamma$ decay
from an EMC shower not associated with a charged track.  
The photon must be in the fiducial region of the EMC,
have a shower-profile consistent with a single photon, and
be well-separated from other showers.  To remove
photons from $\piz$ ($\eta$) decays, we combine the candidate
with other photons having energies of at least
50 (250)\mevcc, and reject it if the invariant mass of any combination
is within 25 (40)\mevcc of the $\piz$ ($\eta$) mass.

We select \Kpm and \pipm candidates from charged tracks consistent with
a kaon or pion mass hypothesis in the
DIRC and in the \dedx  in the SVT and DCH.  We reconstruct \KS
candidates from pairs of oppositely-charged tracks, and determine the
decay vertex with a fit.  We require that the invariant mass
falls within 11\mevcc of the \KS mass; that the distance
between the $B$ decay vertex and the \KS vertex exceeds 5 times the
uncertainty on the distance; and that the angle between the \KS
trajectory and its momentum is less than 100 mrad.  We reconstruct \piz
candidates from pairs of EMC showers each with energy 
$>$50\mev.  We require the invariant mass to be
within 16\mevcc of the \piz mass, and that the energy of each pair in
the \FourS center of mass (CM) frame exceeds 450\mev; this last 
selection is about 83\% efficient.

The dominant source of background is continuum production of light
quark-antiquark pairs, in which a high-energy photon typically is
produced either by initial state radiation, or from the decay of a \piz
or $\eta$ in which one photon is not detected.  To reject these
backgrounds, we construct a Fisher discriminant~\cite{Fisher} from the
polar angle of the $B$ candidate in the CM frame, the angle between the
thrust axis of the $B$ and the thrust axis of the remaining charged and
neutral particles, and the ratio of the second to zeroth
angular moments of the remaining charged and neutral particles around
the thrust axis of the $B$.  We optimize the coefficients independently
in each channel to discriminate between simulated signal and
continuum.

We perform a geometric fit to the reconstructed \B candidate,
with production vertex constrained to the nominal beam spot,
rejecting the candidate if the final state is inconsistent with decay from
a single vertex.  We define $\DeltaE\equiv E^*_B-E^*_\textrm{beam}$ and
$\mes\equiv \sqrt{E^{*2}_\textrm{beam}-\mathbf{p}^{*2}_B}$, where
$E^*_B$ and $\mathbf{p}^*_B$ are the CM energy and momentum of the $B$
candidate, and $E^*_\textrm{beam}$ is the CM energy of each beam.  We
require $\mes>5.2\gevcc$ and $|\DeltaE|<0.15\gev$.  We also require that
the invariant mass of the $K\pi\pi$ system, \mkpipi, fall below
1.8\gevcc; this eliminates much of a rising continuum background 
with very little expected signal loss. It also removes 
$K\pi\pi$ combinations from $D$ decays, in $B\to D\pi^0$ and $B\to D\eta$
where the $\pi^0$ or $\eta$ are mis-reconstructed as a photon.  In an event
in which we reconstruct multiple candidates in one channel that pass the
selection requirements (occurring in 11-27\% of selected signal events, depending
on the channel), we keep the candidate with the largest vertex
probability (with the best \piz mass in the \scn channel; with the \piz
mass as a tie breaker in the \ccn channel) and reject the others.
Candidates reconstructed in different channels are allowed in the same
event.  The dependence of the efficiency of our selection requirements 
on intermediate resonance and on \mkpipi has been checked and 
found to be small; systematic uncertainties are discussed
later.

The dominant backgrounds from \BB events after the selection
criteria have been applied are \btosg processes.  We categorize these
backgrounds: ({\it i\/}) ``crossfeed''
from mis-reconstructed \kppg decays, such as by choosing
incorrectly a particle from the other $B$; ({\it ii\/})
$B\to\kpg$ decays that combine with a track from the other
\B to form a \kppg candidate; and ({\it iii\/}) backgrounds from all
other \btosg decays. A crossfeed candidate may be reconstructed in the
same decay channel in which it is produced, or in a different channel,
and can also be produced in a $B\to\kppg$ decay that is not
used in this analysis (such as $\Bp\to\Kp\piz\piz\gamma$).  We
model our signal as well as crossfeed backgrounds with 
simulated $B\to K_X\gamma$ decays, where $K_X$ is any of the five
lowest-lying $J>0$ kaon resonances above the $K^*(892)$.  We study
backgrounds from \kpg using simulated $B\to K^*(892)\gamma$
and $B\to K_2^*(1430)\gamma$ decays.  We study backgrounds from other
\btosg decays using an inclusive simulation according to the model of
Kagan and Neubert \cite{KN99} with $m_b=4.8\gevcc$, tuned to match
multiplicity distributions measured in inclusive \btosg decays
\cite{semiinclusive}. The largest final background contributions from \btosg
processes are crossfeed backgrounds, for which we obtain yields ranging
from 55\% to 95\% of the signal yields.

We estimate other sources of background candidates from $B$ decays other
than \btosg processes by simulating generic \B decays.  We
pay special attention to \B decays with $K\pi\pi\piz$ and $K\pi\pi\eta$
final states; if the \piz or $\eta$ decays asymmetrically and we don't
detect the lower-energy photon, the kinematic properties of the
resulting \B candidate may resemble a signal candidate.  We
study these decays using high-statistics simulated samples, and look for
signal candidates that are reconstructed from a single $B\to
K\pi\pi\piz$ or $B\to K\pi\pi\eta$ decay.  
We expect to reconstruct fewer than two
such candidates per channel.

We perform a maximum likelihood fit to the joint \mes--\DeltaE
distribution of our selected  candidates.  We fit
all four channels simultaneously to account for crossfeed backgrounds
between channels.  The likelihood function contains terms for correctly
reconstructed signal candidates, crossfeed background candidates between
all 16 combinations of the production and reconstruction channels,
backgrounds from $B\to\kpg$ and from other \btosg decays, and
backgrounds from continuum events.  We have determined from
simulations that the dominant continuum background component adequately
accounts for combinatoric backgrounds from other \BB decays, which do
not show strong peaks in \mes and \DeltaE.

The likelihood function for a candidate reconstructed in decay channel $i$
with kinematic variables $y\equiv(\mes,\DeltaE)$ is given by,
\begin{multline*}
 {\cal L}^i(y)=
  N_{\BB}\left({\cal B}^i \epsilon_s^i f_s^i(y)
               + \sum_j {\cal B}^j \epsilon_x^{ji} f_x^{ji}(y)\right)  \\
  + n_c^i f_c^i(y) + n_b^i f_b^i(y)\ ,
\end{multline*}
where $N_{\BB}$ is the number of \BB pairs in our dataset; ${\cal B}^i$
is the branching fraction for decay channel $i$; $\epsilon_s^i$ and
$f_s^i$ are the efficiency and probability density function (PDF) for
correctly reconstructed signal candidates in decay channel $i$;
$\epsilon_x^{ji}$ and $f_x^{ji}$ are the efficiency and PDF for
crossfeed background candidates produced in channel $j$ and
reconstructed in channel $i$; $n_c^i$ and $f_c^i$ are the yield
and PDF for backgrounds from continuum and generic \BB decay events in
channel $i$; and $n_b^i$ and $f_b^i$ are the yield and PDF for
backgrounds from other \btosg processes in channel $i$.  We
further parameterize the likelihood function by the four data-taking
runs during which data were collected, accounting for slight changes in
experimental conditions.

The branching fractions ${\cal B}^i$, yields $n_c^i$, and shape
parameters of $f_c^i$ are varied in the fit; other
efficiencies, yields, and PDF shapes are fixed from simulation studies.  
We parameterize $f_s^i$ as the product of Crystal Ball
functions~\cite{crystalball} of \mes and of
\DeltaE, $f_x^i$ as the product of a Crystal Ball function of \mes and a
linear function of \DeltaE, and $f_c^i$ as the product of an Argus
function \cite{argus} of \mes and an exponential function of \DeltaE.
We use a binned parameterization for $f_b^i$.  As the signal and
crossfeed terms are both scaled by the parameters ${\cal
B}^i$, the crossfeed background yields vary with the signal
branching fractions, and we measure the branching fractions from
yields of both signal and crossfeed candidates.

\begin{figure}
\includegraphics{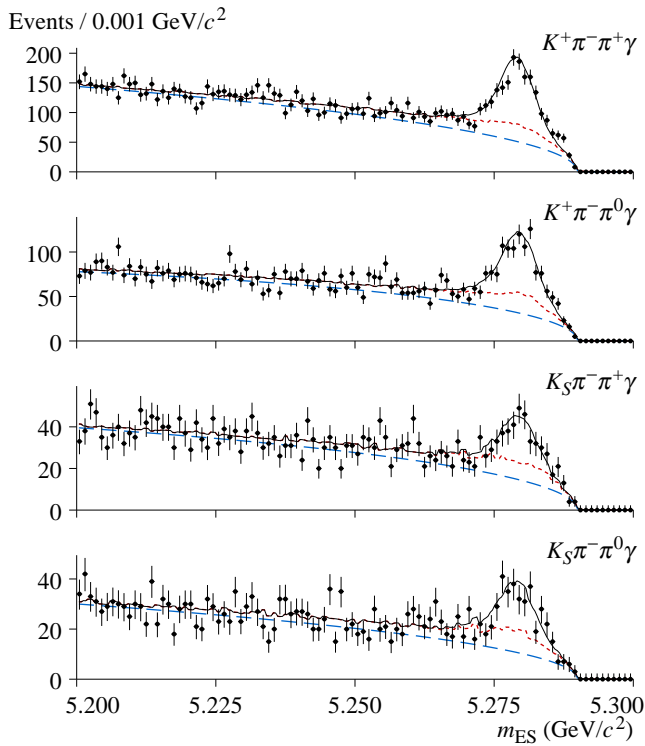}
\caption{Distributions of \mes (points).  Projected \mes\ distributions
from the fit are shown as cumulative curves: continuum
and generic \BB component (dashed), \btosg component (dotted, includes cross-feed), and
signal (solid). The small oscillation in the dotted and solid curves is
 due to the use of binned distributions to model the
 $b\to s\gamma$ component.  }
\label{mes}
\end{figure}

Table~\ref{results} shows the fit results.  Projections in
$\mes$, along with the fit results, are displayed in Fig.~\ref{mes}.
The fit probability ($P$-value) is evaluated with a likelihood 
ratio statistic [10], assuming Poisson-distributed bin contents, to be 10\%. 
The distribution of the test statistic under the null hypothesis 
is evaluated by simulation.

\begin{table}[bp]
\caption{Results of the fit for $B\to K\pi\pi\gamma$, for
$m_{K\pi\pi}<1.8\ \hbox{GeV}/c^2$.  The first error is statistical, the
second systematic.  The yields do not include the channel crossfeeds,
which are included in the fit to obtain the branching fractions.  }
\begin{center}
\begin{tabular}{lcc}  
\hline\hline
\noalign{\vskip1pt}
Channel  & Yield  & Branching Fraction ($10^{-5}$) \\
\hline
\ccc                  &  $899 \pm 38$  & 2.95 $\pm$ 0.13 $\pm$ 0.20 \\
\ccn                  &  $572 \pm 31$  & 4.07 $\pm$ 0.22 $\pm$ 0.31 \\
$K^0\pi^+\pi^-\gamma$ &  $176 \pm 20$  & 1.85 $\pm$ 0.21 $\pm$ 0.12 \\
$K^0\pi^+\pi^0\gamma$ &  $164 \pm 15$  & 4.56 $\pm$ 0.42 $\pm$ 0.31 \\
\hline\hline
\end{tabular}
\end{center}
\label{results}
\end{table} 

\begin{figure}
\includegraphics{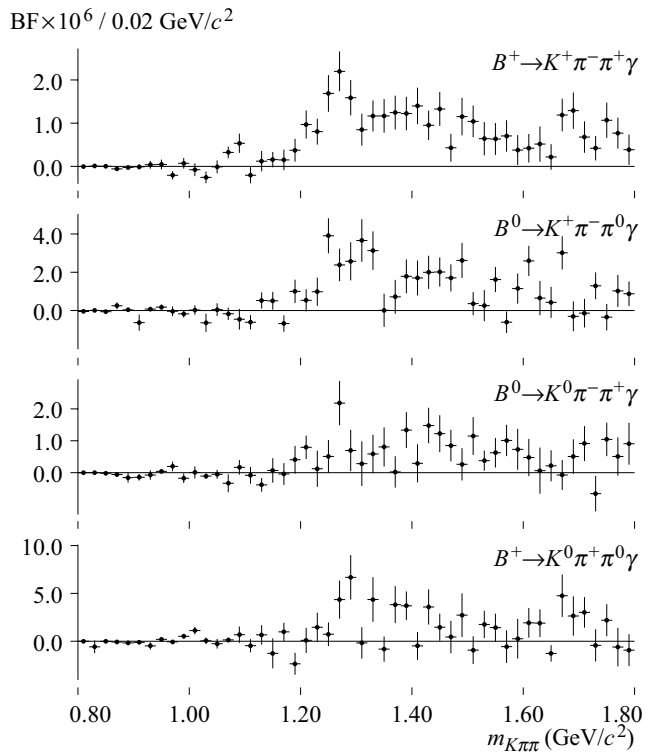}
\caption{Background-subtracted $m_{K\pi\pi}$ spectra.  The branching
fraction in each bin is computed from the weighted event yield.  Error
bars show statistical uncertainties; the systematic uncertainties due to
$b\to s\gamma$ model assumptions are small.}
\label{spectra}
\end{figure}

Figure~\ref{spectra} shows background-subtracted \mkpipi mass
spectra. Background subtraction is achieved using the results of
the fits to calculate event-by-event weights to extract
the signal component~\cite{sPlots}.  We present branching
fractions in bins of \mkpipi, which are largely model-independent,
instead of extracting $B\to K_X\gamma$ branching fractions for specific
$K_X$ resonances.  Disentangling the resonance structure requires
careful modeling of amplitudes and relative phases of
interfering processes, including in the decays of the $K_X$ resonances,
not all of which are well measured.  A partial wave
analysis to extract the resonance structure and measure the photon
polarization should be possible with future datasets.

We validate the procedure for extracting branching fractions and \mkpipi
distributions using fits to
simulated samples.  We verify that the branching fractions
and mass spectra obtained from these toy fits reproduce on average the
simulation inputs.  We use the same procedure to
extract the \mkpipi distributions for continuum and generic \BB
backgrounds and for backgrounds from \btosg decays;
these are consistent with the expected distributions.

\begin{table}[bp]
\caption{Estimated systematic uncertainties in the branching fractions, in percent, by source and
decay channel.}
\begin{center}
\begin{tabular}{lcccc}
\hline\hline
Source & $K^+\pi^-\pi^+$ & $K^+\pi^-\pi^0$ 
       & $K^0_S\pi^-\pi^+$ & $K^0_S\pi^+\pi^0$ \\
\hline
\noalign{\vskip1pt}
\BB count                     &  1.1 &  1.1 &  1.1 &  1.1 \\
\FourS BF                     &  2.6 &  2.6 &  2.6 &  2.6 \\
\hline
Efficiencies:                 &      &      &      &      \\
Photon selection              &  2.7 &  2.7 &  2.7 &  2.7 \\
\piz and $\eta$ veto          &  1.0 &  1.0 &  1.0 &  1.0 \\
Tracking                      &  1.4 &  1.1 &  1.1 &  0.8 \\
Kaon selection                &  4.2 &  4.2 &  1.6 &  1.6 \\
\pipm selection               &  1.4 &  1.0 &  1.4 &  1.0 \\
\piz selection                &      &  3.0 &      &  3.0 \\
Fisher cut                    &  1.0 &  1.0 &  1.0 &  1.0 \\
Vertex probability            &  0.7 &  0.7 &  0.7 &  0.7 \\
MC statistics                 &  0.5 &  0.7 &  0.8 &  1.1 \\
\hline
Backgrounds:                  &      &      &      &      \\
\btosg model                  &  1.4 &  1.0 &  4.0 &  1.3 \\
$B\to K\pi\pi\piz/\eta$       &  0.2 &  0.1 &  0.0 &  0.6 \\
\hline
Beam energy shift             &  1.0 &  0.5 &  1.6 &  0.6 \\
PDF shape                     &  0.1 &  2.9 &  0.9 &  0.2 \\
Fit bias                      &  1.6 &  1.3 &  1.3 &  3.5 \\
\hline
Total                         &  6.7 &  7.6 &  6.7 &  6.8 \\
\hline\hline
\end{tabular}
\end{center}
\label{syserr}
\end{table} 
 
Systematic uncertainties arise from various
sources, shown in Table~\ref{syserr}. The largest
sources are: ({\it i\/}) The \FourS\ branching fractions to
\BpBm and \BzBzb are each assumed to be 0.5. We assign a
2.6\% systematic uncertainty to this, based on current
information~\cite{upsBF}.  ({\it ii\/}) The
uncertainty on the photon selection efficiency
determined from simulated events is estimated to be 2.7\%.  
({\it iii\/}) From studies of $B\to
D\pi^\pm$, $D\to K\pi\pi$ events, we assign an uncertainty of
4.2\% to the charged kaon identification efficiency.  ({\it iv\/}) The
uncertainty of the \piz selection efficiency is estimated at 3.0\%.  ({\it v\/}) There is considerable uncertainty
in the models we use to estimate backgrounds, including 
cross-feed dependence, from \btosg processes.  We
estimate the effect of this uncertainty on both the branching fractions and 
mass spectra by
simulating these backgrounds with substantially
different models.  The largest effect is in the \scc channel, where the
uncertainty is 4.0\%.  ({\it vi\/}) We measure a shift in the
beam energy in $B\to D\pipm$ decays, on average 0.6\mev; we estimate the
effect of this on our fits. ({\it vii\/}) We estimate bias in the fit
due to uncertain parameterization of the signal and background PDFs.
The largest effect is in the \scn channel, where the
uncertainty is 3.5\%.

We have measured branching fractions for $B\to
K\pi\pi\gamma$ in four decay channels for $\mkpipi <
1.8\gevcc$. The $K\pi^+\pi^-$
channels are consistent with the previous measurement~\cite{Belle}. We
present first observations of decays in the $K\pi^+\pi^0\gamma$ channels
that are important to measuring the photon polarization.  The
branching fractions are relatively large in the context of $B\to
X_s\gamma$ decays, providing encouragement that a polarization
measurement may be possible with future datasets.  Mass spectra for the
$K\pi\pi$ system are also presented. We observe an enhancement near
1.3\gevcc and substantial branching fractions at higher
masses. Untangling the resonant contributions presents a challenge for
the polarization measurement.

We are grateful for the excellent luminosity and machine conditions
provided by our \pep2\ colleagues, 
and for the substantial dedicated effort from
the computing organizations that support \babar.
The collaborating institutions wish to thank 
SLAC for its support and kind hospitality. 
This work is supported by
DOE
and NSF (USA),
NSERC (Canada),
CEA and
CNRS-IN2P3
(France),
BMBF and DFG
(Germany),
INFN (Italy),
FOM (The Netherlands),
NFR (Norway),
MES (Russia),
MEC (Spain), and
STFC (United Kingdom). 
Individuals have received support from the
Marie Curie EIF (European Union) and
the A.~P.~Sloan Foundation.




\end{document}